\documentclass[conference]{IEEEtran}
\pdfoutput=1
\usepackage{hyperref}

\usepackage{graphicx}
\usepackage{algorithm}
\usepackage{algpseudocode}
\usepackage{amsmath}
\usepackage{subfigure}
\usepackage{multirow}
\usepackage{booktabs}
\usepackage[numbers]{natbib}

\newtheorem{myDef}{Definition}
\newtheorem{myPro}{Problem}

\begin{document}

\title{Ego-graph Replay based Continual Learning for Misinformation Engagement Prediction}

\makeatletter
\newcommand{\linebreakand}{%
  \end{@IEEEauthorhalign}
  \hfill\mbox{}\par
  \mbox{}\hfill\begin{@IEEEauthorhalign}
}
\makeatother


\author{
\IEEEauthorblockN{Hongbo Bo\IEEEauthorrefmark{1}, Ryan McConville\IEEEauthorrefmark{1}, Jun Hong\IEEEauthorrefmark{2}, Weiru Liu\IEEEauthorrefmark{1}}
\IEEEauthorblockA{\IEEEauthorrefmark{1} University of Bristol, Bristol, UK}
\IEEEauthorblockA{\IEEEauthorrefmark{2} University of the West of England, Bristol, UK} 
\IEEEauthorblockA{\{hongbo.bo, ryan.mcconville, weiru.liu\}@bristol.ac.uk, \{Jun.Hong\}@uwe.ac.uk} 
}

\maketitle

\begin{abstract}
   Online social network platforms have a problem with misinformation. One popular way of addressing this problem is via the use of machine learning based automated misinformation detection systems  to classify if a post is misinformation. Instead of post hoc detection, we propose to predict if a user will engage with misinformation in advance and design an effective graph neural network classifier based on ego-graphs for this task. However, social networks are highly dynamic, reflecting continual changes in user behaviour, as well as the content being posted.  This is problematic for machine learning models which are typically trained on a static training dataset, and can thus become outdated when the social network changes. Inspired by the success of continual learning on such problems, we propose an ego-graphs replay strategy in continual learning (EgoCL) using graph neural networks to effectively address this issue.  We have evaluated the performance of our method on user engagement with misinformation on two Twitter datasets across nineteen misinformation and conspiracy topics. Our experimental results show that our approach EgoCL has better performance in terms of predictive accuracy and computational resources than the state of the art.
\end{abstract}


\section{Introduction}
It is widely believed that the prevalence of misinformation, disinformation and conspiracy theories on online social networks is having profound negative effects on society ~\cite{monti2019fake}.
 Sometimes known as `fake news', dealing with this increasing amount of misinformation is a huge challenge, and numerous automated misinformation detection systems have been proposed. The variety of methods reflects the diversity in the way misinformation manifests online. While some approaches focus solely on the content, for example, the text of a tweet on Twitter ~\cite{buntain2017automatically}, others adapt approaches where the social context of the misinformation is taken into account ~\cite{dou2021user, bian2020rumor, monti2019fake} and misinformation is predicted based on the network structure. However, most systems are concerned with the prediction of whether an existing \textit{post} online is misinformation. Instead, we take a new approach, where we seek to predict whether a \textit{user} will engage with misinformation (e.g., post, retweet, like, comment). This is a foundational step in developing systems that not only can detect misinformation, but also intervene.

 Graph machine learning approaches involving online social networks typically represent users as nodes in the graph, with edges representing certain types of relationships between users.  A graph is usually represented as an adjacency matrix in which each element indicates whether the node pair is connected.
  Typical tasks using this representation include  node classification ~\cite{kipf2017semi, velivckovic2017graph} and link prediction ~\cite{kipf2016variational}. Recently, graph neural networks (GNNs)~\cite{scarselli2008graph} have been shown to be effective in such graph machine learning tasks. 

 Ego-graphs, also called ego-networks, are an alternative representation, where nodes are composed of a single central node (ego) and the neighbours of that node. In general, ego-graphs focus more on the relevant user instead of studying the network as a whole, instead of studying the network as a whole. For tasks where the objective is to predict the behaviour of a user, this is an appealing formulation. Thus, in this work, we combine ego-graphs with graph neural networks in our approach to demonstrate how it can effectively predict whether a user will engage with misinformation based on their own ego-graph. 

 However, as online social networks are continuously changing by, for example, reacting to major events (e.g., elections, pandemics), the performance of a static model is expected to decay over time.  At the same time, the social network structure and attributes can also change rapidly in a short period of time.  
In terms of graph representation of social networks, this is reflected with simple changes such as the addition or removal of edges and nodes, but potentially also with more complex high-level changes, such as changes to node attributes.
 A naive approach is to simply retrain the model on an updated training dataset but this has several drawbacks, which are often addressed by online learning~\cite{fontenla2013online}. In online learning, the model can be updated incrementally with each new data point without having to (re-)train on the entire dataset at once.  
 However, without care, deep neural network based online learners tend to suffer from a problem known as catastrophic forgetting when learning new tasks, where previous knowledge is forgotten in the process of learning new knowledge.

Continual learning~\cite{kirkpatrick2017overcoming, wang2020streaming, zhou2020continual} has been proposed as an approach to obtain models which can continuously adapt to new data and tasks, without suffering from catastrophic forgetting. The aim of continual learning is to learn tasks sequentially, with two general goals: (1) learning a new task without leading to catastrophic forgetting of former tasks, (2) leveraging knowledge from prior tasks to facilitate the learning of new tasks.
As a category of continual learning methods, experience replay has proven effective in many research areas~\cite{zhou2020continual, lopez2017gradient}

The existing experience replay based on graph neural networks~\cite{zhou2020continual, wang2020streaming} have focused on the node replay strategy, which does not fully capture the neighbourhood information. In order to overcome this limitation, we first propose an ego-graph based replay strategy in graph neural networks for continual learning, called ego-graph replay, which can provide sufficient neighborhood information during replay.
We then demonstrate the effectiveness of our proposed approach for misinformation engagement prediction in dynamic social networks. We show that our proposed novel ego-graph based experience replay strategy  in continual learning is able to adapt to different types of misinformation and accurately predict user engagement with nineteen misinformation and conspiracy theories.

\paragraph{Contributions} Firstly, we provide a new formulation of the problem to facilitate future interventions on misinformation where the objective is to predict whether users will engage with misinformation. Secondly, using this formulation, we demonstrate how an ego-graph based graph neural network can accurately predict whether users will engage with nineteen categories of misinformation and conspiracy theories across two different datasets. Further, we reveal how this model could suffer from catastrophic forgetting when exposed to the dynamic nature of online social networks and misinformation. Finally, to address this we propose a novel ego-graph based experience replay approach using GNNs (EgoCL) that can effectively address this issue, and show how our novel approach out-performs the state of the art on two different Twitter datasets.

\section{Related Work} 
\paragraph{Graph Neural Networks}
Graph neural networks (GNNs) have rapidly grown to become a popular research area and have been used in supervised and unsupervised tasks, such as node and graph classification, graph generation and clustering. For the node classification problem, supervised GNNs have shown impressive levels of performance, such as GCN~\cite{kipf2017semi} and GAT~\cite{velivckovic2017graph}.

\paragraph{Social Networks and Ego-graphs}
Since ego-graphs (ego-networks) can provide locality-sensitive information, they are widely used in social network analysis, such as community detection ~\cite{shchur2019overlapping} and social influence prediction~\cite{qiu2018deepinf,bo2020social, bo2021social}. Some of the research focuses on the ego-graph itself. In \cite{arnaboldi2017online} the ego-graph is layered into different circles and used for information diffusion. In \cite{arnaboldi2013ego} differences in sizes and structures of the Twitter ego-networks are analysed when the users have strong connections.

\paragraph{Continual Learning}
 One line of continual learning focuses on regularization-based strategies. The regularization is introduced to maintain the stability of the model parameters that make contributions to the previous tasks, so that the trained model does not forget knowledge from previous tasks when learning new tasks. Examples include EWC~\cite{kirkpatrick2017overcoming} and GCL~\cite{tang2020graph}. Another line of research focuses on experience replay strategies~\cite{rolnick2018experience}. Replay-based methods use limited data from the previous tasks or have a generator to simulate data from the previous tasks as part of the input to the current task, such as ER-GNN~\cite{zhou2020continual} and GEM~\cite{lopez2017gradient}.

\paragraph{Misinformation Detection}
 Misinformation detection on social networks has already attracted a lot of attention in the research community. The key difference in our research is that we propose to predict whether a user will engage with misinformation, while previous research mostly focuses on the detection of misinformation itself. In the misinformation detection work, graph structures have been used successfully in BiGCN~\cite{bian2020rumor} and GCNFN~\cite{monti2019fake}, with both using the GCNs to detect the misinformation spreading patterns. In UPFD~\cite{dou2021user}, various user preferences have been captured simultaneously by joint content and graph modelling.  Continual learning has also been used in misinformation detection by GNN-CL~\cite{han2020graph} where they utilize the GEM~\cite{lopez2017gradient} and EWC~\cite{kirkpatrick2017overcoming} to detect `fake news' in a continual learning setting. 

\section{Predicting User Engagement with Misinformation}
In this section, we first formulate our problem of predicting if a user will engage with misinformation.  We then describe how social network data can be processed such that it can be used within this formulation. Finally, we will describe our proposed ego-graph based graph neural network classifier  for this problem.

\subsection{Problem Definition}
Let $G=(V,E)$ be a graph which consists of a set of nodes $V$ and a set of edges $E$,  where $E\in V\times V$. A social network can be represented by a graph $G$, where $V$ represents users and $E$ is the set of edges representing how the users are connected. As an input to the graph neural network, a graph can also be represented as $G=(A,X)$, where $A_{|V|\times|V|}\in\{0,1\}$ is the adjacency matrix and $X_{|V|\times d}$ is the $d$-dimensional matrix of node features which can be customized for specific problems. 

Users in social networks perform social actions, such as posting a tweet, retweeting an existing tweet. We call a \textit{social action} that results in the user acting on misinformation a \textit{misinformation action}.

\begin{myDef}{\textbf{Misinformation Actor:}}
A user $v$ is called a \textbf{misinformation actor} when the user has performed a misinformation action; otherwise $v$ is called a non-misinformation actor. $s_v\in \{0,1\}$ is the label of user $v$, where $s_v=1$ indicates the user has performed the misinformation action and $s_v=0$ indicates otherwise.
\end{myDef}
Given a social network, we aim to build a classifier that is able to predict whether each user in the network will engage in a misinformation action and hence will be identified as a misinformation actor.


\begin{myPro}{\textbf{Misinformation Engagement Prediction:}}
Given a social network $G$ for a specific type of misinformation, the misinformation engagement prediction problem is formulated as a binary node classification problem which predicts $S=\{s_v: v\in V \}$ by learning a classifier on $G$.
\end{myPro}

\subsection{Data Processing} \label{Data}
In this subsection, we introduce how we process two datasets into a set of social networks related to different types of misinformation. 

Given the dynamic nature of a social network and misinformation on the network, instead of using a single graph representing the entire social network, we extract a set of social networks related to different types of misinformation and conspiracy theories. This way, we are able to analyze and evaluate whether the properties of these networks are different, and if a model trained on one network performs well on others.

\subsubsection{Hashtag Dataset}
For the first dataset, we collected 540,042 raw tweets from Twitter. Given the prevalence of misinformation around COVID-19, we extract tweets mentioning COVID-19 and a number of hashtags related to different types of misinformation and conspiracy theories.  The list of hashtags used is shown in Table \ref{stat1}. 

In order to build a social network (and then ego-graphs), we extracted the \textit{mention} relationships from these tweets. On Twitter, a mention is a specific action a user can take where they reference other Twitter users in their tweets.
A tweet could be an original post, a retweet, or a retweet with a quote. If there exists such a relationship between two users in a tweet, we add an edge between this pair of users in the graph.

Given the collected data including ten types of misinformation and conspiracy theories (we call them ten misinformation topics for short), we extract a sub-graph for each misinformation topic to study misinformation engagement prediction under each misinformation topic.

 As each hashtag sub-graph has heavy class imbalance towards misinformation in that sub-graph (i.e., in the network around each topic, many users are \textbf{misinformation actors}), we augment each misinformation hashtag sub-graph  with 10,000 additional mention relationships from the much broader COVID-19 mention graph in order to make the task more realistic.
For each of the hashtags, the 10,000 additional mention relationships have no overlap with the existing mention  relationships under the hashtag. We choose a certain number of additional relationships instead of a certain proportion as this ensures that the proportion of positive labels for each hashtag is as different as possible to reflect more realistic online social networks.

Secondly, for each user in the graph we also extract the user's \textit{follower count, following count, tweet count and listed count} as node features. For each node, we assign a label 1 or 0 depending on whether the user posted, retweeted (including with a comment), or replied with a hashtag related to misinformation. 
The statistics of the 10 extracted graphs are shown in the Table~\ref{stat1}.

\subsubsection{MuMiN Dataset}
 In the first dataset, we identified tweets related to different types of misinformation using the known hashtags, that while useful, is a rather coarse measure. In order to address this, we use a recent dataset, known as the MuMiN dataset~\cite{NielsenMcConville2022}\footnote{We used the version \textbf{0.1.4}}, which connects tweets to fact-checked claims on a broad range of topics. Using this dataset, we process the data in a similar way, with one significant difference. In this dataset we aim to predict not only if a user is a misinformation actor, but also if they are a misinformation actor within a specific topic of misinformation. Therefore, for each topic, we will add additional users into the graph who have engaged with other misinformation topics outside of those included in the tasks. This allows us to study not only if we can detect misinformation actors, but also if we can detect misinformation actors for a specific misinformation topic.
 We end up with a further nine misinformation topics to study, which are shown in Table ~\ref{stat2} and we will refer to our processed version as MuMiN-mentions.

\begin{table}[h]
\centering
\caption{The statistics of the Hashtag dataset. The columns represent the names of the hashtags, numbers of nodes, numbers of links, average degree and percentage of positive labels.}

\begin{tabular}{l|l|l|l|l}
\hline
Hashtags & Nodes & Links & Degree & Pos. Labels($\%$)  \\ \hline
plandemic      & 31,464 & 42,609 & 2.708       & 62.89              \\ \hline
darktolight    & 8,207  & 12,125 &2.954        & 9.18                 \\ \hline
wwg1wga        & 20,118 & 37,680 & 3.745       & 36.25                  \\ \hline
scamdemic      & 13,787 & 18,030 & 2.615       & 20.73                \\ \hline
greatreset     & 14,366 & 18,989 & 2.643       & 27.32                   \\ \hline
thegreatreset  & 12,207 & 17,655 & 2.892       & 22.14                   \\ \hline
greatawakening & 10,964 & 14,385 & 2.624       & 18.25                   \\ \hline
agenda2030     & 31,604 & 62,552 & 3.958       & 33.37                  \\ \hline
agenda21       & 8,292  & 13,090 &3.157        & 9.03                    \\ \hline
chinesevirus   & 19,781 & 32,341 & 3.269       & 34.89                   \\ \hline
\end{tabular}
\label{stat1}
\end{table}

\begin{table}[h]
\centering
\caption{The statistics of the MuMiN-mentions graphs: The columns are the same as in Table~\ref{stat1}, except that the first column represents topics.}

\begin{tabular}{l|l|l|l|l}
\hline
Topics                 & Nodes & Links & Degree & Pos. Labels($\%$)  \\ \hline
violence police   &11,179   &15,963  & 2.855  &3.92                \\ \hline
military reports  &13,257   &18,421   & 2.779        &3.18                     \\ \hline
narendra modi  &7,290   &13,238   & 3.631        &3.38                   \\ \hline
reduce population  &16,197   &22,033   & 2.720        &3.00                \\ \hline
fires australia &20,195   &26,954   &2.669        &3.08                \\ \hline
biden said &15,695   &20,828   &2.654         &3.61                  \\ \hline
 wisconsin voted &10267   &13749   &2.678         &5.19                   \\ \hline
martian sunset &11366   &13715   & 2.413        &2.74                   \\ \hline
syrian air defense &23682   &32651   &2.76        &3.55                   \\ \hline
\end{tabular}
\label{stat2}
\end{table}

\subsection{Misinformation Engagement Prediction} \label{non-cl}
In this section we describe the proposed system, and its variants that are evaluated, for the problem of predicting misinformation engagement on social networks.
\subsubsection{Graph Neural Network}
We will first describe the GNN used to predict misinformation engagement.

\paragraph{Input layer}
The input layer of the GNN has a feature vector for each node representing a 64-dimensional pre-trained DeepWalk embedding with additional node features, specifically, \textit{following count}, \textit{follower count}, \textit{tweet count} and \textit{list count} (\textit{list count} was not available in  the MuMiN dataset).

\paragraph{GNN layer(s)}
While there is no restriction on the GNNs architecture we could use, we will experiment with a common approach. A GNN model is defined as a function $f(X,A)$, where $X$ is the node feature matrix of the adjacency matrix $A$.  For the node classification task, we will use a graph attention network (GAT)~\cite{velivckovic2017graph}.  The ${(l+1)}^{th}$ hidden GNN layer $H$ can be defined as:
\begin{equation}
    H^{(l+1)}=\sigma(\widetilde A H^{(l)}W^{(l)}),
\label{gnn}\end{equation}
where $\sigma(\cdot)$ denotes an activation function, $W^{(l)}$ is the weight matrix and $\widetilde{A}$ is the adjacency matrix that defines the aggregation strategy from neighbors. The input layer $H^{(0)}=X$. In our experiments we use a GAT and the element $\alpha_{ij} \in \widetilde{A}$ in a GAT layer is computed as:
\begin{equation}
     e_{ij}=attn(W^{(l)}h_i,W^{(l)}h_j),
\end{equation}
\begin{equation}
    \alpha_{ij}=softmax_j(e_{ij})={\textstyle\frac{Exp(e_{ij})}{\sum_{k\in N_i}Exp(e_{ik})}},
\end{equation}
where $h_i,h_j\in H^{(l)}$ and the attention function $attn$ is instantiated with a dot product and a LeakyReLU~\cite{xu2015empirical} non-linearity.

\paragraph{Output layer}
The output layer will be activated by the log-softmax function which will output a two-dimensional vector for each user. We compare the vectors of the users with the ground truth set $Y$, and then optimize the negative log-likelihood loss.

\subsubsection{Ego-Graphs}

Inspired by DeepInf~\cite{qiu2018deepinf}, we extract ego-graphs from each graph, but experiment with two different ego-graph extraction strategies. An ego-graph~\cite{zimmermann2008predicting} is typically the graph of all nodes that are within a certain distance from a node, which means ego-graphs have the same radius length but different graph sizes. However, for GNN-based mini-batch training, it is more convenient if input graphs are of the same size, so we propose two extraction strategies for building fixed-size ego-graphs.

\paragraph{Breadth-first search (BFS)}Breadth-first search (BFS) is a search strategy which starts from a node in the graph, finds all the adjacent nodes to that node, and then finds all the adjacent nodes of each adjacent node in turn, and so on. For each social network $G$, we use BFS to extract the ego-graph of each node with a fixed ego-graph size $n$, instead of a fixed distance. In this way, we can ensure the nearest $n-1$ nodes to the central node are added to the ego-graph.
\paragraph{Random walk with restart (RWR)} Random walk with restart (RWR) randomly samples several different paths from an initial node through a network with a fixed walk length. We use RWR to extract the ego-graph of each node and ensure each ego-graph will have the same radius length, but we note that this does not ensure the nearest nodes have been added.

If during the BFS or RWR ego-graph extraction process we cannot find enough neighbour nodes to build ego-graphs, we add some dummy nodes (to make up to ego-graph size) and add 0 to indicate relationships between the ego node and these dummy nodes.

\subsection{Evaluation}
We use the Hashtag dataset and MuMiN-mentions dataset described in Section~\ref{Data} to evaluate the proposed ego-graph based approach to classify whether users are misinformation actors. For comparison, we compare this approach with a standard GAT, which does not make use of ego-graphs. The results are shown in Table~\ref{rslt1} and Table~\ref{rslt2} . For both the GAT and ego-graphs based GATs, we divide the dataset into 75\%, 12.5\%, and 12.5\% splits for training, validation and test sets respectively. We use two layer GATs with each layer containing 128 hidden units with 8 attention heads. The learning rate is 0.01 and the number of training epochs is set as 100. For both BFS and RWR, the ego-graph size is set as 50.

As shown in Table~\ref{rslt1}, we can see the ego-graph GAT model with the BFS strategy has the best performance on most types of misinformation in the Hashtag dataset where it achieves better performance in 7 of 10 hashtags.
In the second dataset, as shown in Table ~\ref{rslt2}, we can see that ego-graph based methods exhibit excellent results compared to a regular GAT. In the MuMiN dataset we are addressing a more challenging problem, not only the prediction of whether a user is involved in misinformation, but also if they are involved in a specific type of misinformation. In this task, Ego-BFS again works very well.

\begin{table}[h]
\centering
\caption{The performance of GATs framework and ego-graphs GATs based framework on Hashtag dataset.}
\begin{tabular}{l|l|l|l}
\hline
Hashtags               & GAT       & Ego-RWR       & Ego-BFS \\ \hline
scamdemic      &0.7578     & 0.7659    &\textbf{0.7951}   \\ \hline
wwg1wga        &0.8173     & 0.8829    &\textbf{0.8893}     \\ \hline
agenda21       &0.7492     & 0.6435    &\textbf{0.7911}     \\ \hline
greatawakening &0.7331     & 0.6526    &\textbf{0.7668}     \\ \hline
plandemic      &0.8635     & 0.9050     &\textbf{0.9161}     \\ \hline
thegreatreset  &0\textbf{.7979}     & 0.7414    &0.7838     \\ \hline
greatreset     &\textbf{0.7557}     & 0.6880     &0.7287     \\ \hline
chinesevirus   &0.764      & 0.8219    &\textbf{0.8361}     \\ \hline
darktolight    &\textbf{0.8414}     & 0.6799    &0.8321     \\ \hline
agenda2030     &0.7413     & 0.8036    &\textbf{0.8292}    \\ \hline
\end{tabular}
\label{rslt1}
\end{table}

\begin{table}[h]
\centering
\caption{The performance of GATs framework and ego-graphs GATs based framework on MuMiN-mentions dataset.}

\begin{tabular}{l|l|l|l}
\hline
Topics               & GAT       & Ego-RWR       & Ego-BFS \\ \hline
violence police       &0.6198     &0.7663    &\textbf{0.8426}  \\ \hline
military reports       &0.6667     &0.7267      &\textbf{0.8783}    \\ \hline
narendra modi      &0.6967     &0.8151      &\textbf{0.9825}   \\ \hline
reduce population      &0.6422    & 0.7336           &\textbf{0.8524}     \\ \hline
fires australia      &  0.6668     &  0.8810      &\textbf{0.9289 }    \\ \hline
biden said  &  0.6371    &\textbf{0.8845}     &0.8802   \\ \hline
wisconsin voted     & 0.6188     &0.7763     &\textbf{0.8433 }    \\ \hline
martian sunset   &0.6269     &\textbf{0.9523 }    &  0.9304     \\ \hline
syrian air defense   & 0.6250      &  0.8575 & \textbf{0.9085}   \\ \hline
\end{tabular}
\label{rslt2}
\end{table}

\section{Continual Misinformation Engagement Prediction}
\label{section:DI}
While these results are promising and show the effectiveness of an ego-graph based approach for the task, the static setting in which we evaluated the systems are not reflective of the previously discussed dynamic and evolving nature of social networks and misinformation topics. 
In order to verify this, we will simulate the emergence of different misinformation topics, incrementally training each of the deep neural networks on the topics as they emerge.

We adapt the Hashtag and MuMiN-mentions datasets for this setting.  We first create a series of graphs $G=\{G^1,G^2,...,G^N\}$ corresponding to a series of tasks $T=\{T^1,T^2,...,T^N\}$ which are encountered sequentially. The prediction problem is defined as:
\begin{myPro}{\textbf{Incremental Misinformation Engagement Prediction}}
For each task $T^i$, where $i=1,2,..., N $, we have the social network $G^i$. Misinformation engagement prediction is a binary node classification problem which aims to predict $S^i=\{s^i_v: v\in V^i \}$ by learning a classifier $C^i$ by incrementally training $C^{i-1}$ on $G^i$.
\end{myPro}

When a model has been trained on task $T^i$, we will use all of the previous tasks from $T^1$ to $T^{i-1}$ as the test sets, and measure the Area Under Curve (AUC).
Specifically, we are interested in understanding if the model is able to retain performance on previous misinformation topics (tasks) as it is trained on a new topic of misinformation.
We calculate the average AUC over 10 runs and show the results in Figure~\ref{node_forget}.
In the figure, `fully retrained' describes a model retrained on all of the data so far. The `incrementally trained' model is trained in an online manner, without retraining on previous data.
From this it is clear that the model that is incrementally trained suffers from a drop in performance as soon as it is incrementally trained on a new task. The same problem exists for the proposed ego-graph approaches.

\begin{figure}[h]
    \centering
    \includegraphics[width=0.9\linewidth]{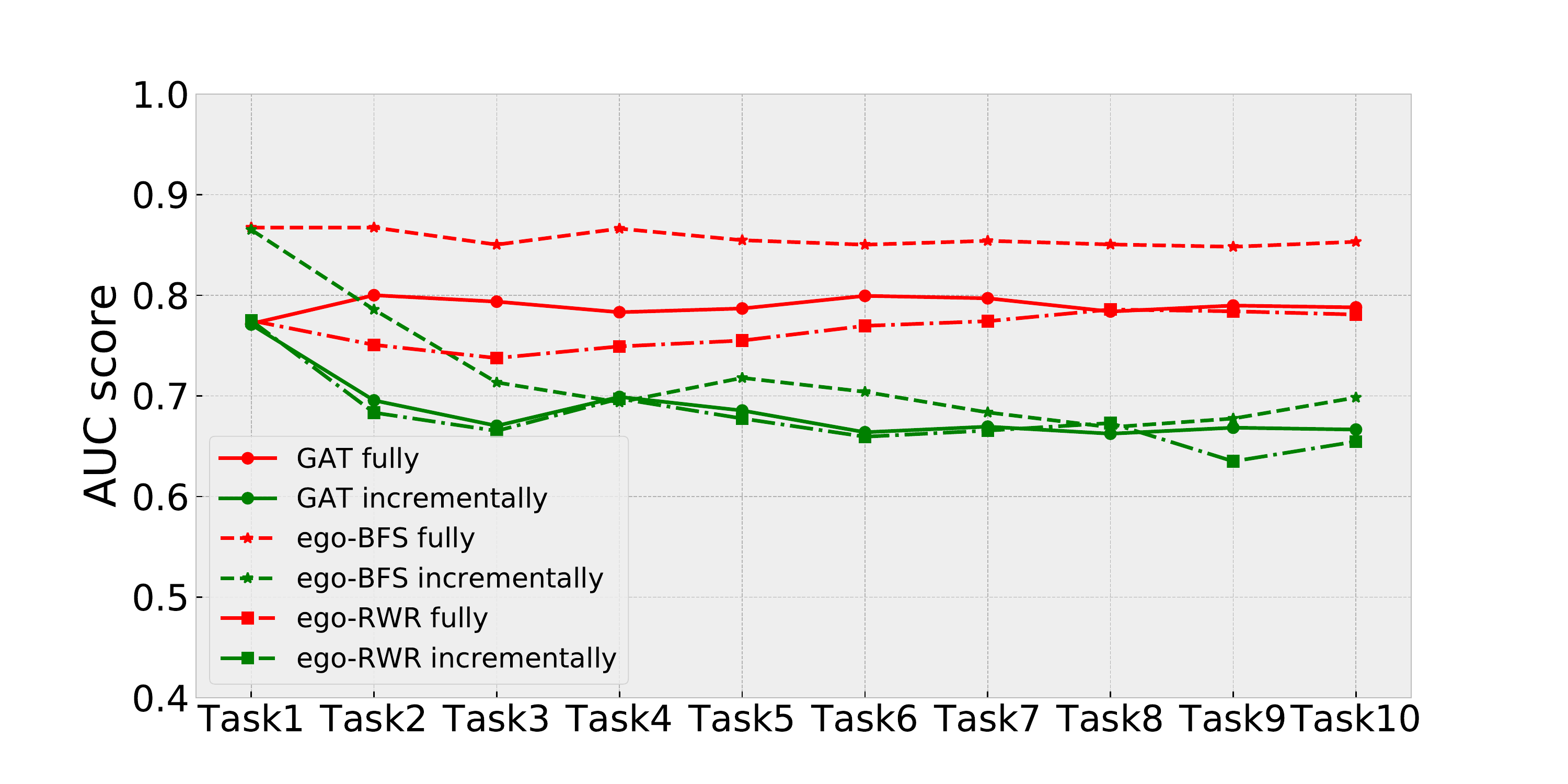}
    \caption{The catastrophic forgetting in GAT, ego-graphs (BFS) GAT, and ego-graphs (RWR) GAT on Hashtag dataset.  The incrementally trained models are in green while fully retrained models in red.  }
    \label{node_forget}
\end{figure}

This is an example of catastrophic forgetting~\cite{mccloskey1989catastrophic} which refers to the tendency that a model `forgets' previously learned knowledge upon learning new knowledge. The most intuitive cause of catastrophic forgetting is that model parameters tend to adapt towards fitting new data from a new task and thus deviate from optimal values of previously learned tasks. A common approach to deal with this issue is \textit{continual learning}.

In the setting of continual learning, the objective is to continually train a classifier that can effectively learn a series of tasks $T$ instead of only an individual task typical in traditional settings. For each new task $T^i$, the input data contains not only the data for the new task but also some replay data from the previous tasks. For example, the misinformation content in a previous task may be the \textit{\#chinesevirus}, whilst the current task could be about \textit{\#thegreatreset}. Due to the inherent constraint in continual learning that once the learning of a task is completed, the full dataset from this task is no longer available, methods need some measure to ensure that what is subsequently learned does not negatively impact what was previously learned.

\subsection{Ego-graphs replay: EgoCL} \label{section:ER}
In order to deal with the catastrophic forgetting in continual misinformation engagement prediction, we propose our novel framework \textbf{EgoCL} based on the ego-graph replay strategy. The framework is shown in Figure~\ref{fig:fw}. 

\begin{figure}[h]
    \centering
    \includegraphics[width=0.5\textwidth]{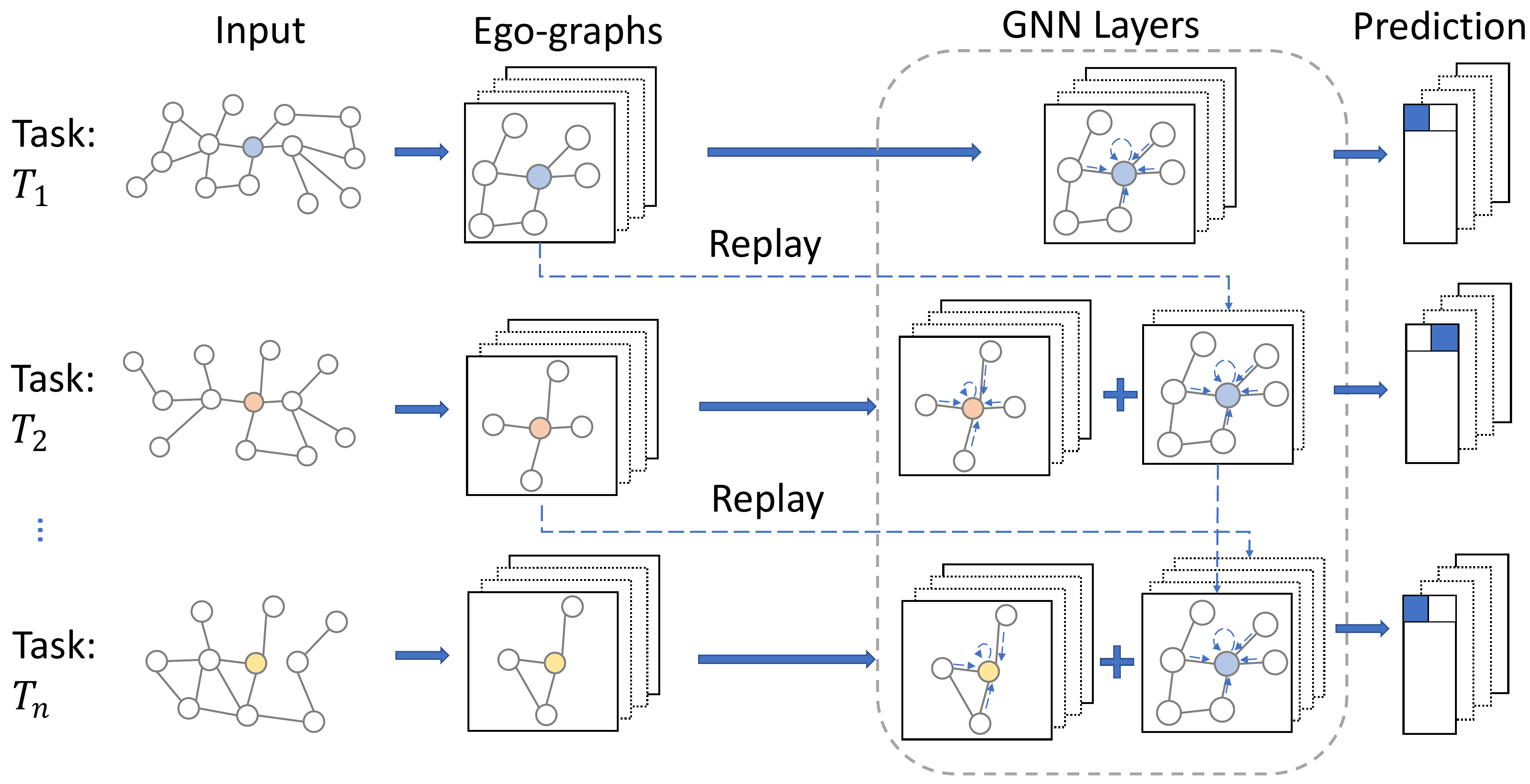} 
    \caption{The EgoCL framework. For each task, ego-graphs will be extracted from the entire input graph and then combined with the replay samples from the previous tasks as input to the GNNs. 
    }
    \label{fig:fw}
\end{figure}

\paragraph{Ego-graph replay}

\begin{algorithm}[h]
\caption{EgoCL}

\label{al:Al2}
\begin{algorithmic}[1]
\Require
    The processed graph $G^i$ of each continual task $T^i$ with features; sample rate $r$; ego-graph size $n$.
\Ensure
  Model which can mitigate catastrophic forgetting of preceding tasks.
\State Initial empty replay set $R_s$;
\State Initial GNN model;
\State Train:
    \For {task $T^i$, $i=1...N$} 
        \State Sample the size $n$ ego-graphs $G^i_{ego}$ from $G^i$
        \State Train model by the combined set $G^i_{ego} \cup R_s$ ;
        \State Randomly select samples $R_s^i$ from $G^i_{ego}$ according to $r*|G^i_{ego}|$; 
        \State $R_s=R_s\cup R_s^i $;
    \EndFor
\end{algorithmic}
\label{cl}
\end{algorithm}

For the current learning task, the input data will be extended to include sample ego-graphs extracted from the previous tasks randomly.
The process is shown in Algorithm~\ref{al:Al2}:

\subsection{Experimental Evaluation}
\subsubsection{Experiment Setup}\label{setup}
Our experimental evaluation aims to evaluate two important aspects of our proposed method. 
We would first like to measure the effectiveness of the proposed continual learning strategy by comparing its performance with the non-continual learning models discussed in Section~\ref{non-cl}. Specifically, we will compare with a graph attention network (GAT), BFS ego-graph based GAT (Ego-BFS-GAT) and RWR ego-graph based GAT (Ego-RWR-GAT).

Secondly, we would also like to compare the effectiveness of the proposed ego-graphs based experience replay with other previously proposed continual learning strategies.
\begin{itemize}
    \item Node replay: Random node replay is in general used as a baseline in graph continual learning research~\cite{zhou2020continual,wang2020streaming}. We introduce the experience replay strategy into the implementation of the GATs framework with a random node selection scheme. For each task $T^i$, we randomly select $r*|G^i|$ replay nodes, where $r$ is the sample rate and $|G^i|$ is the number of nodes in graph $G^i$.
    \item EWC: Elastic weight consolidation~\cite{kirkpatrick2017overcoming} is an importance-based weight regularization continual learning method that can reduce catastrophic forgetting by strengthening the restriction for the important parameters in previous tasks.
    \item ER-GAT-MF: This ~\cite{zhou2020continual} is a continual learning GNN with an experience replay-based method which replays nodes closest to the average feature vector calculated by `Mean of Feature'. In our experiments, we use a GAT as the GNN layers in ER-GNN-MF.
\end{itemize}

Thirdly, to demonstrate the scalability of EgoCL with other GNNs, we tested EgoCL by using GCN~\cite{kipf2017semi} as the classification network.

The performance is evaluated with two metrics, the average AUC score (Equation \ref{eq:auc}) as well as the average forgetting (FGT) of each model (Equation \ref{eq:fgt}). The average AUC score measures the model classification performance in which the higher the AUC is, the better the model's performance.  The average FGT measures the catastrophic forgetting phenomenon of the model and the lower the FGT is, the better the model's ability in dealing with forgetting. 
\begin{equation}
    \frac1N\sum_{i=1}^NAUC_{N,i}
    \label{eq:auc}
\end{equation}
\begin{equation}
   \frac1{N-1}\sum_{i=1}^{N-1}(AUC_{i,i}-AUC_{N,i})
        \label{eq:fgt}
\end{equation}
where $AUC_{i,j}$ denotes the test AUC score on task $T^j$ after the model has finished task $T^i$ training.

The baseline methods and our methods are all based on 2 layer GAT models where each layer contains 128 hidden units with 8 attention heads. The  learning rate is 0.01 with $1e^{-3}$ decay rate and trained for 100 epochs. For all replay-based methods, the replay rate $r$ (in algorithm~\ref{al:Al2}) is set as 0.1. The ego-graph size is 50.

\subsubsection{Experimental Results}
\begin{table*}[h]
\centering
\caption{The performance of two EgoCL models on two datasets, along with the performance of the baselines. }
\begin{tabular}{l|l|l|l|l|l}
\hline
\multicolumn{2}{l|}{\multirow{2}{*}{{Methods}}} & \multicolumn{2}{l|}{Hashtags}                             & \multicolumn{2}{l}{MuMin-mentions} \\ \cline{3-6} 
\multicolumn{2}{l|}{}                           & AUC                  & FGT(\%)            & AUC                     & FGT(\%)         \\  \bottomrule   
                         
\multirow{3}{*}{Non-CL}     &GAT                & 0.6580$\pm$0.0263    & 13.68$\pm$1.90      & 0.5949$\pm$0.0362        &4.41$\pm$2.03            \\\cline{2-6} 
                            & Ego-BFS-GAT     & 0.6593$\pm$0.0211    & 17.89$\pm$6.31     & 0.7837$\pm$0.0171         &6.37$\pm$2.10           \\\cline{2-6}  
                            & Ego-RWR-GAT      & 0.6453$\pm$0.0743    & 17.24$\pm$8.13     & 0.6801$\pm$0.0386         &7.42$\pm$4.07          \\ \bottomrule

\multirow{5}{*}{CL}         & EgoCL-BFS-GAT           &0.7924$\pm$0.0109 &2.61$\pm$1.79      & 0.8109$\pm$0.0232   & 2.01$\pm$1.32    \\\cline{2-6} 
                            &EgoCL-RWR-GAT           & 0.7471$\pm$0.0221   & 2.86$\pm$1.34     &  0.7390$\pm$0.0212   & 5.23$\pm$1.79    \\ \cline{2-6} 
                            & Node Replay       & 0.6774$\pm$0.0158    & 9.90$\pm$3.14    & 0.6106$\pm$0.0399     & 2.90$\pm$1.89        \\ \cline{2-6} 
                            & EWC               & 0.6229$\pm$0.0522    & 8.69$\pm$2.01    & 0.6134$\pm$0.0319      & 3.45$\pm$1.91    \\ \cline{2-6} 
                            & ER-GAT-MF         & 0.6533$\pm$0.0526  & 9.83$\pm$2.61      & 0.7342$\pm$0.0302  &6.85$\pm$2.70   \\ \bottomrule 
\multirow{2}{*}{Other GNN-based EgoCL} &EgoCL-BFS-GCN &\textbf{0.7993}$\pm$0.0091 &\textbf{2.28}$\pm$1.40  &\textbf{0.8239}$\pm$0.0413 & 2.67$\pm$1.14 \\ \cline{2-6} 
                                 &EgoCL-RWR-GCN & 0.7539$\pm$0.033& 5.88$\pm$3.17 & 0.7632$\pm$0.0112 &  4.21$\pm$1.76 \\ \hline
\end{tabular}
\label{baselines}
\end{table*}

The experimental results are shown in Table~\ref{baselines}.
For both datasets and tasks, our proposed continual learning ego-graph neural networks,  EgoCL-BFS and EgoCL-RWR, show better performance over most of the baselines in term of AUC. Further, when we compare our two proposed methods, EgoCL-RWR performs slightly worse than EgoCL-BFS. We think that this is because the nearest neighbours of a user provide the most information for this task, which BFS specifically captures, while RWR does not.

Comparing the performance of GCN- and GAT-based EgoCL methods, GCN-based achieves the best results, but GAT-based EgoCL is very close and not much worse. GCN computes its hidden representation by taking an unweighted average over its neighbours’ representations and the attention mechanism of GAT does not play an effect. This indicates that under a certain misinformation contexts, for an ego-node, there is no single neighbour node that is particularly important.

\subsubsection{Hyper-parameters Analysis}
We also analyze the  hyper-parameters to test the robustness of our proposed methods. We compare with the values of the hyper-parameters mentioned in Section~\ref{setup} unless stated otherwise,and we use GATs as the GNN-classifier.

\paragraph{Replay rate}
We use different replay rates of 0.01, 0.05, 0.1, 0.2 and 0.3 to evaluate the performance of our proposed ego-graph based experience replay. The results are shown in Figure~\ref{fig:4}. We can observe a slow increase in  AUC and decrease of the average FGT performance when we use a  higher replay rate. This is expected as when we increase the replay rate, we are replaying more data from previous tasks during the current task. A replay rate of 1 would be equivalent to fully retraining the model at each task. However, we can see that with low replay rates it is possible to achieve a high AUC and low forgetting rate, with higher replay rates resulting in only marginal improvements.
\begin{figure}[htb]

	\centering
	\subfigure[EgoCL-BFS-GAT on the Hashtag dataset.]{
		\begin{minipage}[b]{0.45\linewidth}
			\includegraphics[width=1\linewidth]{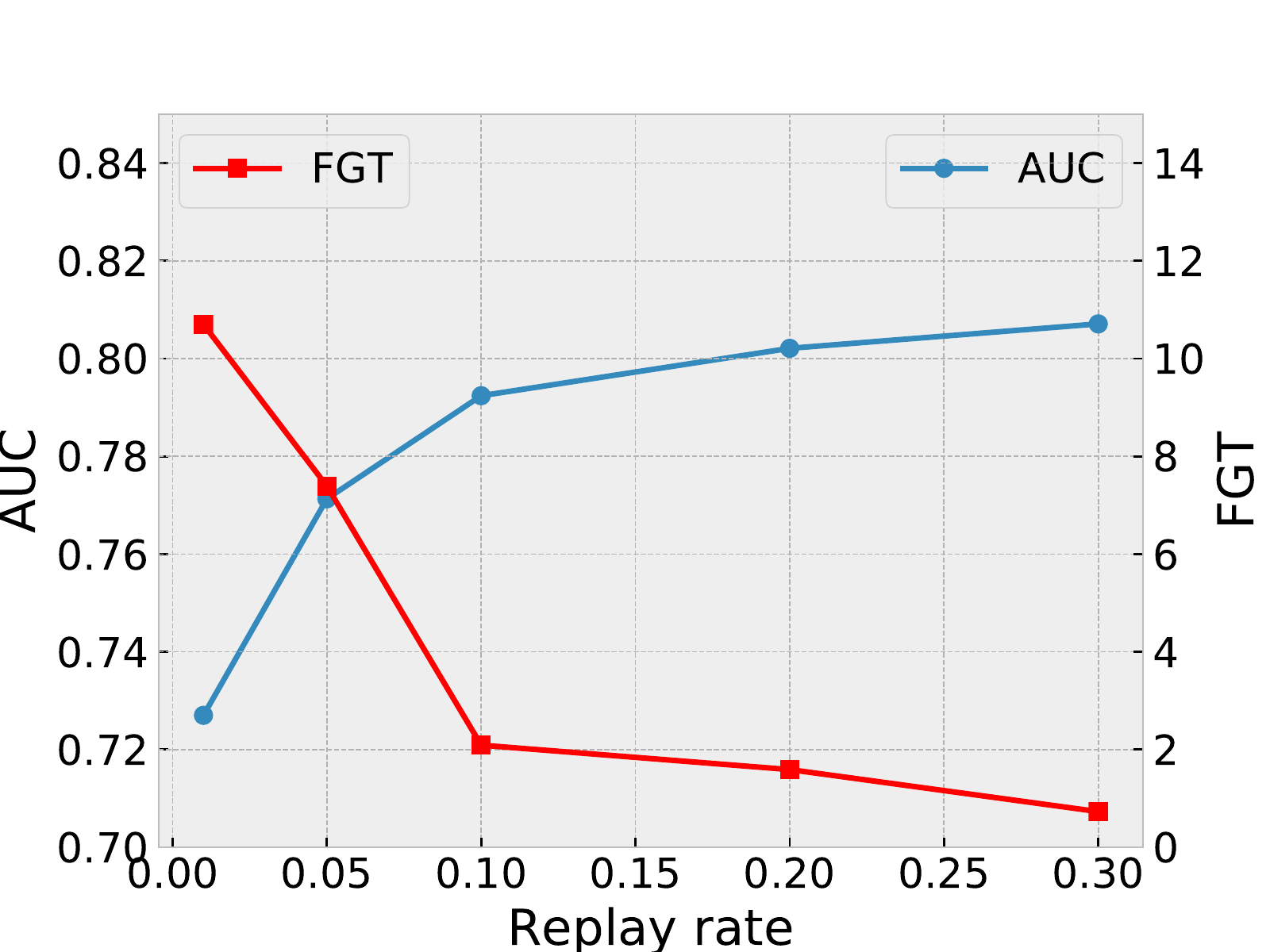} 
		\end{minipage}
		\label{fig:4.1}
	}
    	\subfigure[EgoCL-RWR-GAT on the Hashtag dataset.]{
    		\begin{minipage}[b]{0.45\linewidth}
   		 	\includegraphics[width=1\linewidth]{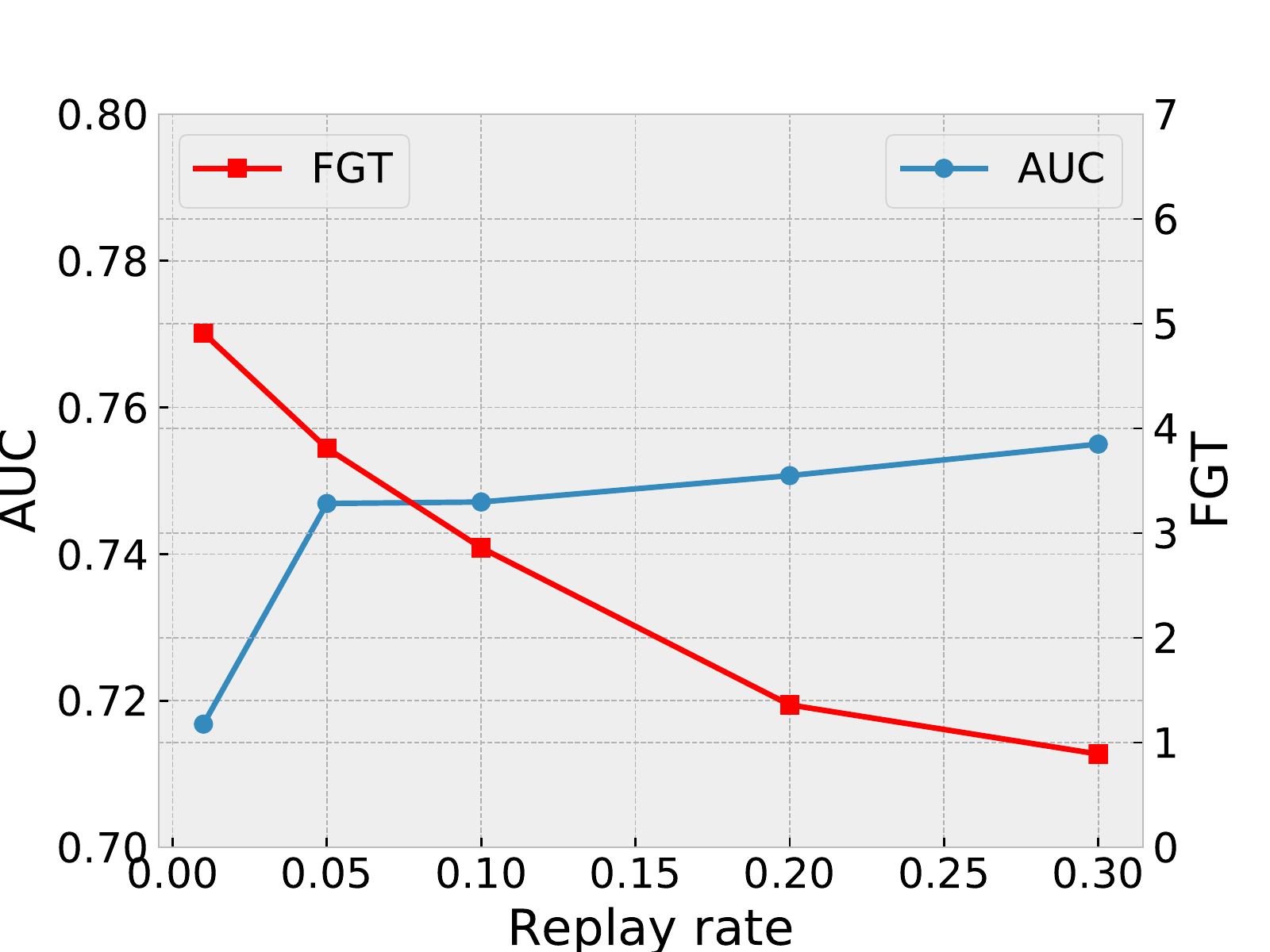}
    		\end{minipage}
		\label{fig:4.2}
    	}
	\\ 
	\subfigure[EgoCL-BFS-GAT on the MuMiN-mentions dataset.]{
		\begin{minipage}[b]{0.45\linewidth}
			\includegraphics[width=1\linewidth]{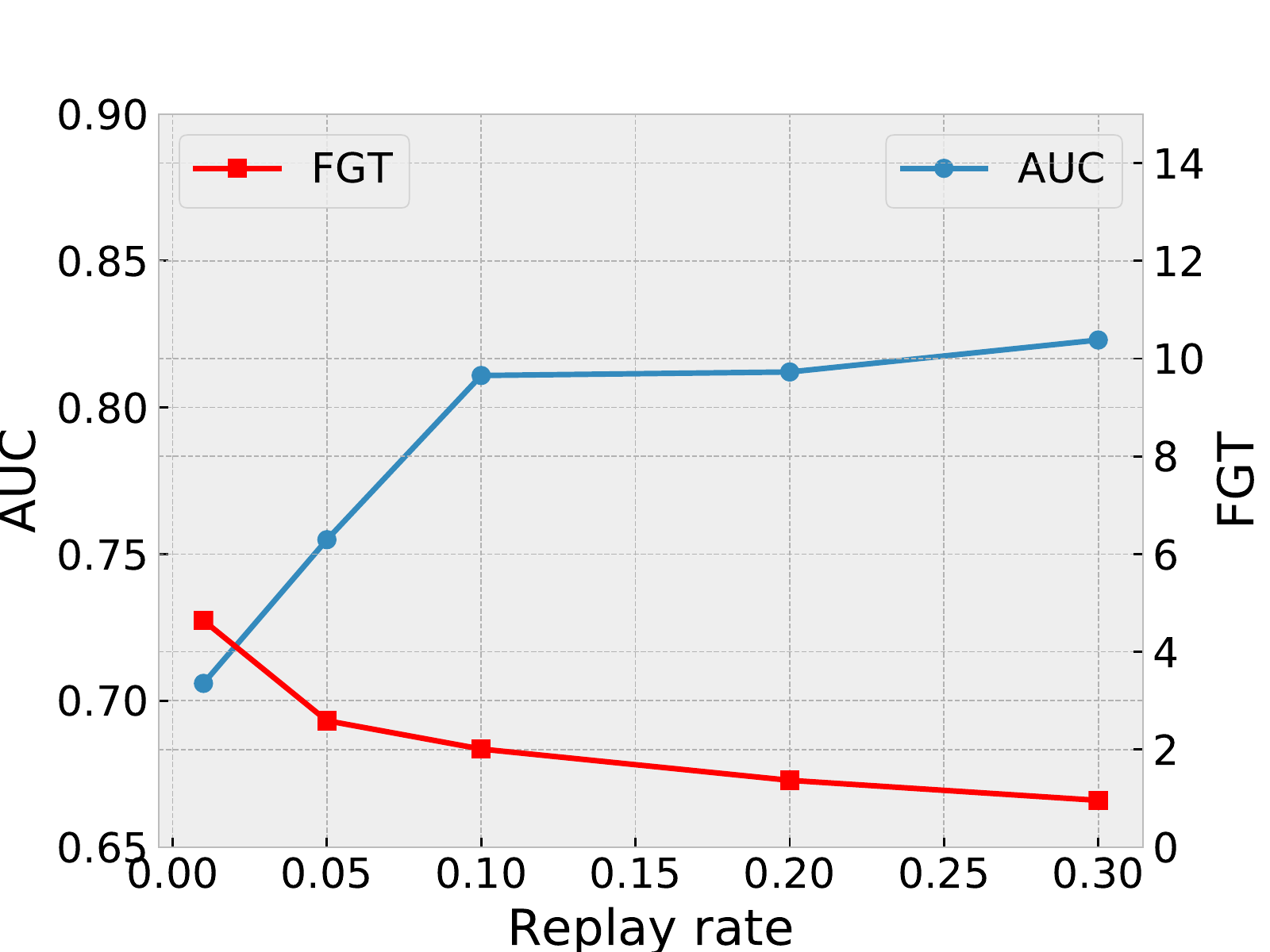} 
		\end{minipage}
		\label{fig:4.3}
	}
    	\subfigure[EgoCL-RWR-GAT on the MuMiN-mentions dataset.]{
    		\begin{minipage}[b]{0.45\linewidth}
		 	\includegraphics[width=1\linewidth]{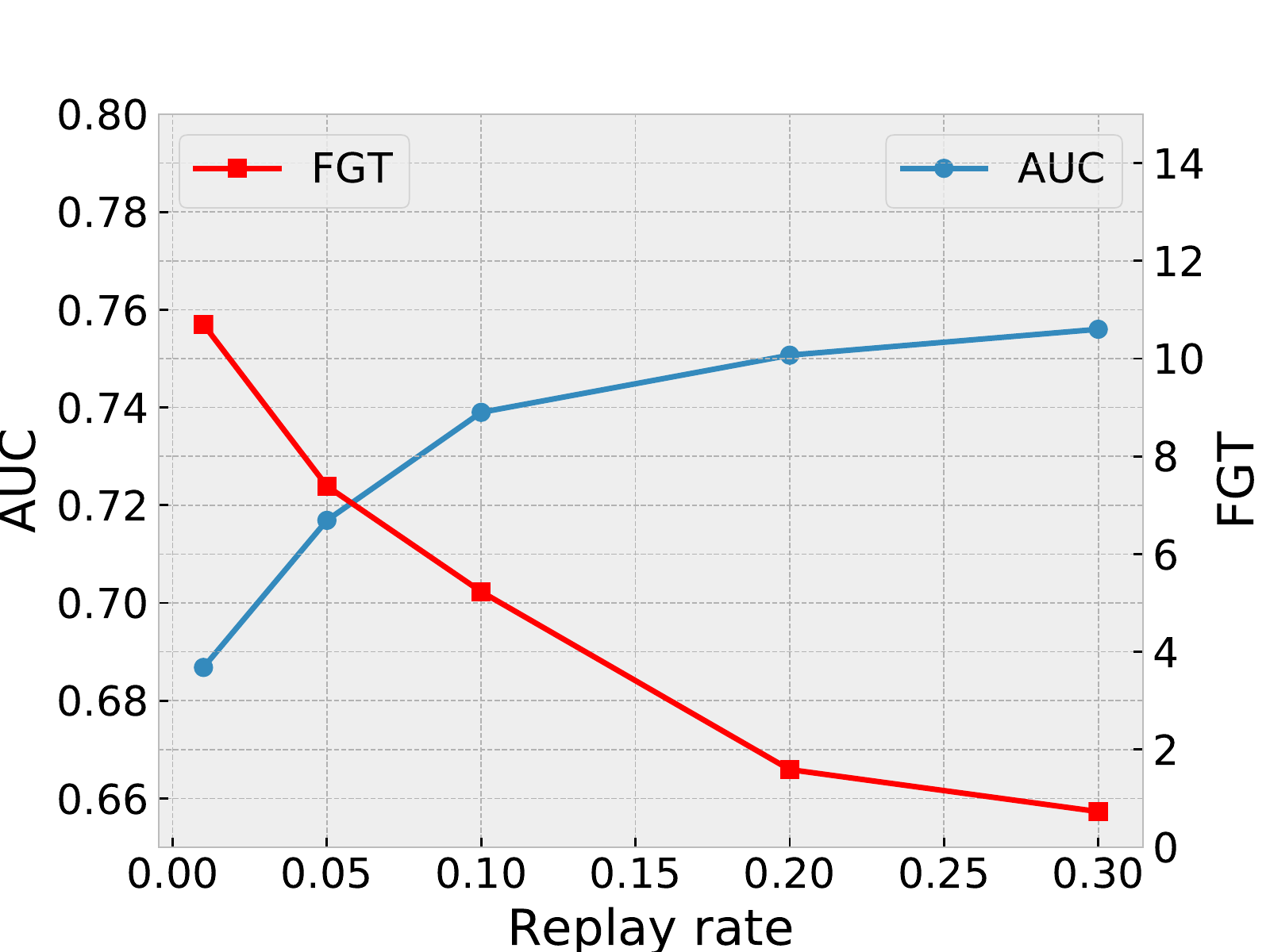}
    		\end{minipage}
		\label{fig:4.4}
    	}
	\caption{Replay rate analysis}
	\label{fig:4}
\end{figure}
\begin{table}[]
\centering
\caption{Resource comparison of the continual learning methods. The values of each column have been normalized with the lowest value set as $1.00$.  Replay data size is denoted as $R$ and model parameter size denoted as $M$.}

\begin{tabular}{l|l|l|l|l|l}
\hline
\multirow{2}{*}{} & \multicolumn{2}{l|}{GPU Memory} & \multicolumn{2}{l|}{Computation} & \multirow{2}{*}{\begin{tabular}[c]{@{}l@{}}Additional \\ storage\end{tabular}} \\ \cline{2-5}
                  & train         & test        & train         & test         &                                                                         \\ \hline
EgoCL             & 1.00          & 1.00        & 2.89          & 1.00         & $M+R$                                                                     \\ \hline
Node Replay       & 4.11          & 3.38        & 1.00          & 3.11         & $M+R$                                                                   \\ \hline
ER-GAT-MF            & 1.00          & 1.00        & 18.27         & 1.04         & $M+R$                                                                 \\ \hline
EWC               & 3.34          & 3.33        & 2.55          & 46.15        & $2*M$                                                                   \\ \hline
\end{tabular}
\label{cost}
\end{table}
\paragraph{Ego-graph size} 
We use RWR to extract different sizes of ego-graphs (from 20 nodes to 80 nodes) to evaluate the performance on both datasets and the results are shown in Figure~\ref{fig:size1}. We can see that when the size of the ego-graph increases, the average AUC scores of EgoCL-BFS are not increased much on both datasets. We believe that this is because the mention graph is typically sparse, and 20-node ego-graphs capturing the nearest neighbours of a node can already provide enough information as to whether the user will engage with misinformation. 
We note, however, that the performance of 80 node ego-graphs obtained by RWR sampling is worse than that of 20-node ego-graphs obtained by BFS sampling, indicating that local structure captured by BFS is more useful.
\begin{figure}[htb]

	     \subfigure[EgoCL-BFS on the Hashtag dataset.]{
		\begin{minipage}[b]{0.45\linewidth}
			\includegraphics[width=1\linewidth]{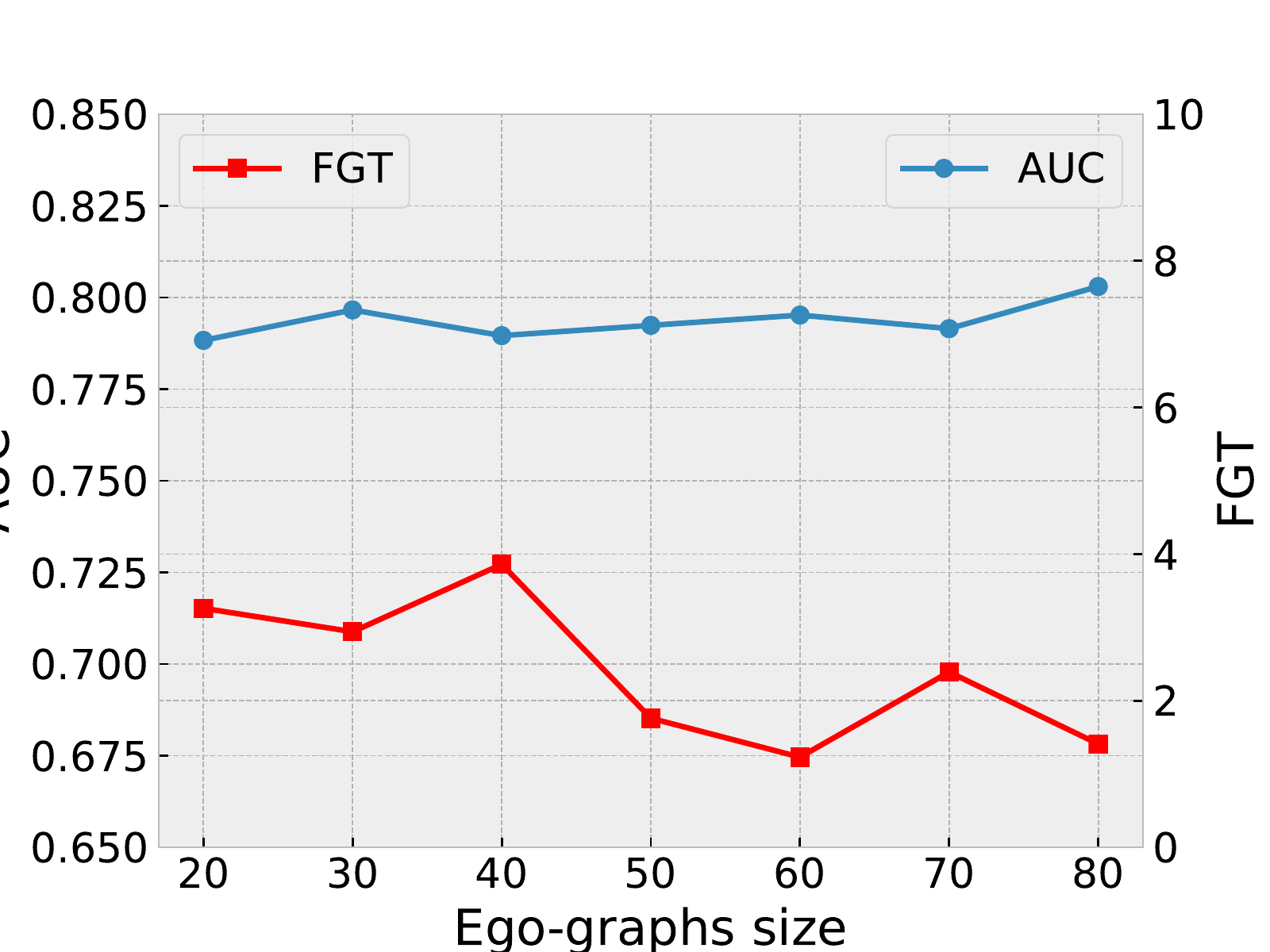} 
		\end{minipage}

		}
	    \subfigure[EgoCL-RWR on the Hashtag dataset.]{
		\begin{minipage}[b]{0.45\linewidth}
			\includegraphics[width=1\linewidth]{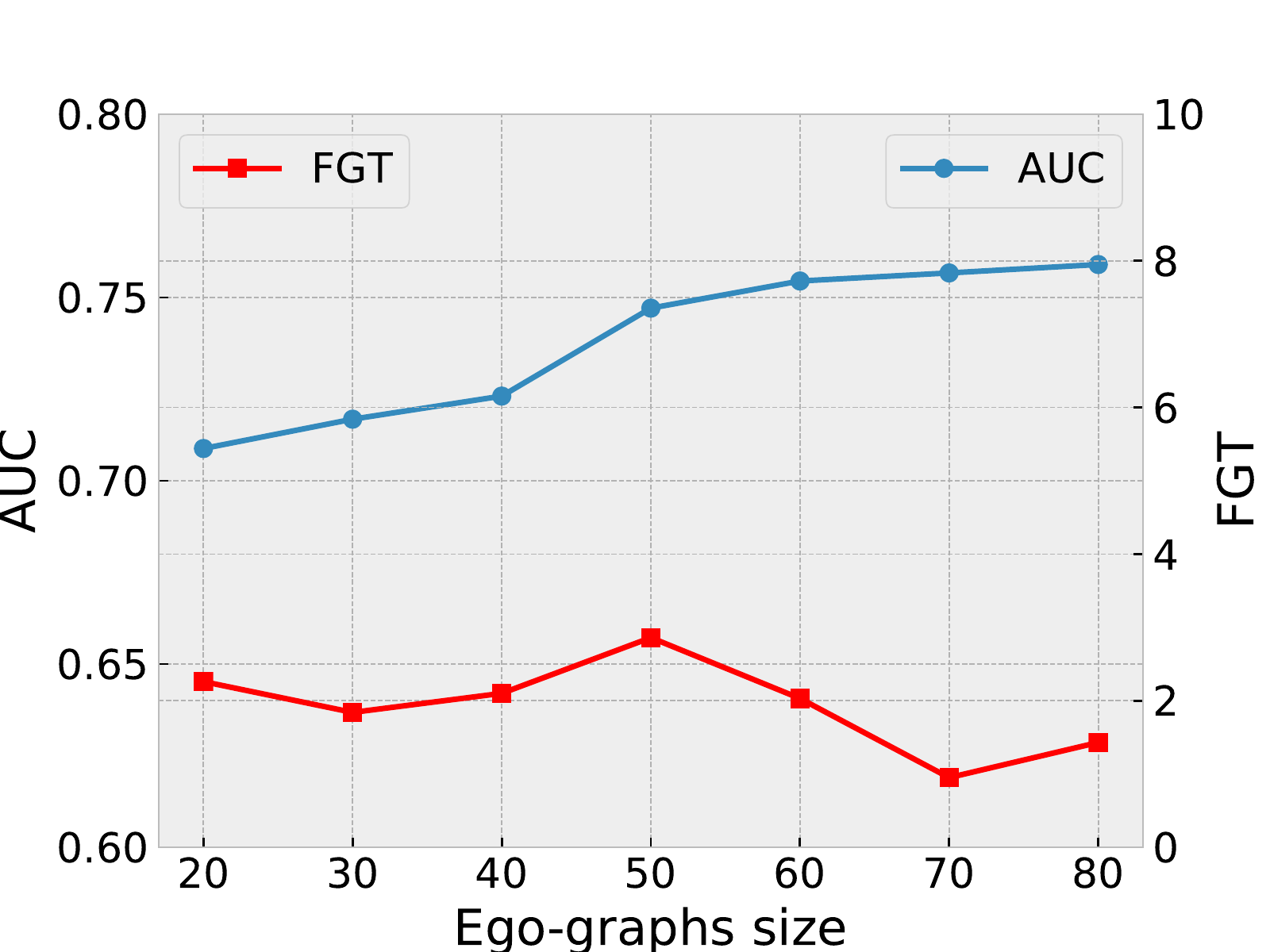} 
		\end{minipage}

		}
	\\
		\centering
	\subfigure[EgoCL-BFS on the MuMiN-mentions dataset.]{
		\begin{minipage}[b]{0.45\linewidth}
			\includegraphics[width=1\linewidth]{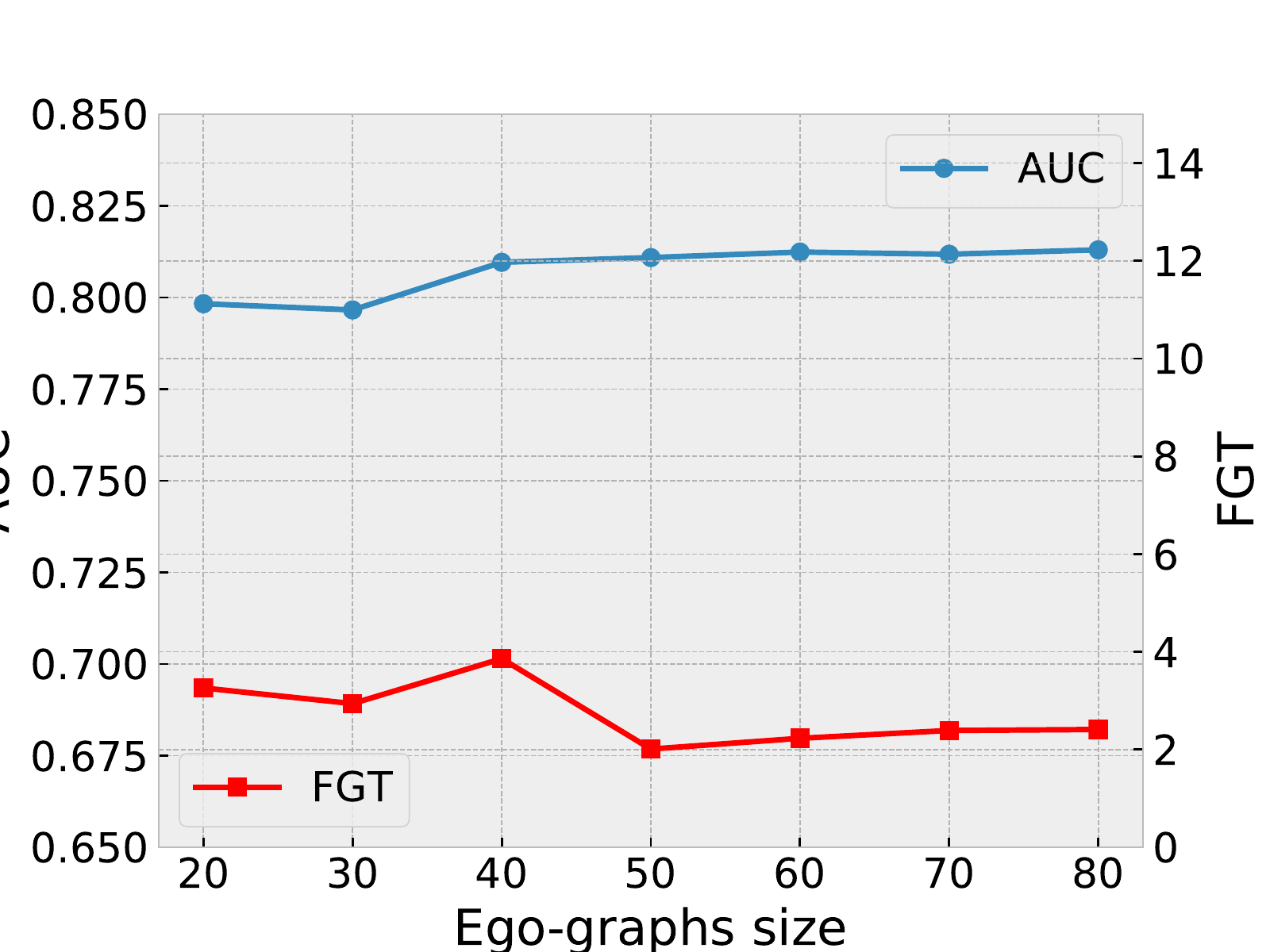} 
		\end{minipage}

	}
    	\subfigure[EgoCL-RWR on the MuMiN-mentions dataset.]{
    		\begin{minipage}[b]{0.45\linewidth}
   		 	\includegraphics[width=1\linewidth]{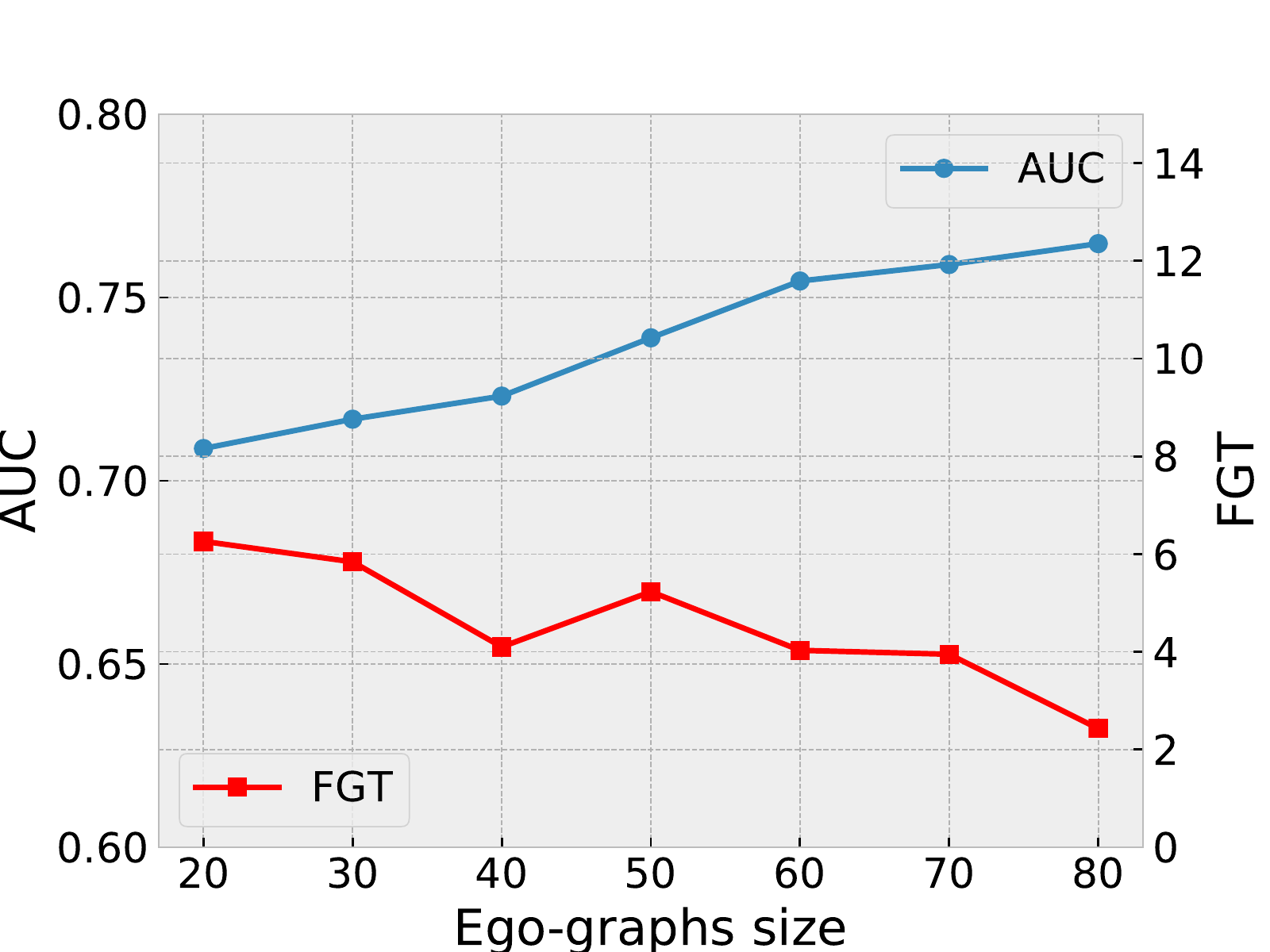}
    		\end{minipage}

    	}
	\caption{Ego-graph size analysis}
	\label{fig:size1}
\end{figure}

\subsubsection{Resource comparison}
We also visualize relative GPU memory requirements and computation time for each of the methods in Table~\ref{cost}. Due to the fact that EgoCL and ER-GAT-MF both use mini-batch training on ego-graphs, the GPU memory costs are low. 
Node replay and EWC both require the entire graph as the model input and as such have larger GPU memory requirements. In terms of  computation time, our proposed EgoCL approach has the second longest training time but the shortest test time. 

\subsubsection{Timestamp setting}
Temporality can be an important consideration when dealing with misinformation. We redesign five tasks of the MuMiN-mention dataset based on timestamp information.
In previous experiments, the tasks were designed to be misinformation content-related; each task of the MuMiN-mention dataset had a different topic and the topic subgraph of each task has been sampled into two different ego-graphs sets (BFS and RWR) which further used for our approach. However, in the current temporality testing, the new tasks have been designed to incorporate notions of temporality and sequences of nodes in the ego-network engaging with each other over time, which means each task is time-related. We blended the instances of topics in the MuMiN-mention dataset and reconstructed five mention-subgraphs based on the posting time of each instance-related tweet as our five tasks: We slice a time period\footnote{The time period of each task: Task 1: 2018.05.25-2019.08.28; Task 2: 2019.08.28-2020.02.25; Task 3: 2020.02.25-2020.05.06; task 4: 2020.05.06-2020.07.30; Task 5: 2020.07.30-2020.09.28.} for each task and take the users who post or reply to tweets in this time period as the nodes of the mentioned subgraph combining with the mentioned relationship between those users to construct a subgraph. These five time-related subgraphs are then used for subsequent operations such as ego-graph sampling and comparative experiments with baselines.
We compare our method EgoCL with other continual learning methods and the results are shown in Table~\ref{temporality}. From the test results, EgoCL-BFS-GAT has the best performance. The EgoCL-BFS methods can still make accurate predictions on social networks that change over time. At the same time, GAT-based has a lower forgetting rate than GCN-based EgoCL, which indicates that EgoCL-GAT produces less forgetting in time-sequential misinformation engagement prediction tasks. However, different from the previous experimental results, the inferiority of EgoCL-GCN is demonstrated in this set of experiments. We need more experiments to further analyze the reasons for this situation, which will be our follow-up work.
\begin{table}[]
\centering
\caption{Temporality testing: The hyper-parameters are the same as evaluation setting in~\ref{setup}}
\begin{tabular}{l|l|l}
\hline
              & \multicolumn{2}{l}{MuMiN-mentions}  \\ \hline
              & AUC    & FGT   \\ \hline
EgoCL-BFS-GAT & \textbf{0.8859}$\pm$0.0089 & 4.89$\pm$1.12  \\ \hline
EgoCL-RWR-GAT & 0.7076$\pm$0.0132 & \textbf{4.21}$\pm$1.29  \\ \hline
EgoCL-BFS-GCN & 0.8239$\pm$0.0996 & 12.58$\pm$6.32 \\ \hline
EgoCL-RWR-GCN & 0.7234$\pm$0.0221 & 12.02$\pm$7.71 \\ \hline
Node Replay   & 0.6324$\pm$0.0688 & 6.33$\pm$3.26  \\ \hline
EWC           & 0.6518$\pm$0.0329 & 5.12$\pm$3.81  \\ \hline
ER-GAT-MF     & 0.7012$\pm$0.0210 & 6.24$\pm$2.73  \\ \hline
\end{tabular}
\label{temporality}
\end{table}

\section{Conclusions}
In this paper, we have proposed a method for predicting whether users will engage with misinformation and conspiracy theories on Twitter, and if so, what types of misinformation or conspiracy theory they will engage with. The proposed method utilizes ego-graphs with a graph attention network, formulated as a binary node classification task. We demonstrate the superiority of an ego-graph approach on two Twitter datasets, one using hashtags and another using fact-checked claims. 
Furthermore, we show how this style of approach is not well suited for the type of online learning required for online social networks where  the network naturally changes over time, as well as having the dynamics of misinformation topics, with the models suffering from catastrophic forgetting. 
To address this we propose a novel continual learning based approach, building on the aforementioned ego-graph neural network and show how it is able to continually learn to predict if users will engage with misinformation as new conspiracy theories and topics arise, while addressing catastrophic forgetting.
 We compare our proposed ego-graph replay based continual learning approach with the state-of-the-art on two different Twitter datasets and tasks and empirically show that our proposed method, EgoCL, has better performance in terms of predictive accuracy and computational resources than the state of the art.

\section*{Acknowledgements}
We would like to thank Dan Saattrup Nielsen for sharing and helping with the MuMiN dataset.

\footnotesize
\bibliographystyle{IEEEtran}
\bibliography{reference}

\end{document}